\documentclass[epj]{svjour}

\usepackage{graphicx}
\usepackage{fancyhdr}
\def\makeheadbox{{
 }}
\input paperdef
\def\mytitle{My title} 
\def\myauthors{My name}  
\def\mytype{My type of session}
\def\mysession{My session}

\def\mytitle{SUSY in the Light of 
$B$ Physics and Electroweak Precision Observables} 
\def\myauthors{Georg Weiglein}    
\def\mytype{Contributed Talk}    
\def\mysession{Colliders - SUSY Phenomenology}

\begin{document}
\title{SUSY in the Light of \boldmath{$B$} Physics and Electroweak 
Precision Observables}
\author{G.~Weiglein\inst{1}
}                     
%
%
\institute{IPPP, University of Durham, Durham DH1 3LE, U.K.}
%
\date{}
\abstract{
Indirect information about the possible scale of
supersymmetry (SUSY) breaking can be obtained from the comparison 
of precisely measured observables (and also of exclusion limits)
with accurate theory predictions
incorporating SUSY loop corrections. 
Recent results are reviewed obtained from a combined analysis of
the most sensitive electroweak precision observables (EWPO),
$\MW$, $\sweff$, $\Ga_Z$, $(g-2)_\mu$ and $\Mh$, and
$B$-physics observables (BPO),
$\br(b \to s \ga)$, $\br(B_s \to \mu^+\mu^-)$,
$\br(B_u \to \tau \nu_\tau)$ and $\De M_{B_s}$.
Assuming that the lightest supersymmetric particle (LSP) provides the
cold dark matter density preferred by WMAP and other cosmological data,
$\chi^2$~fits are performed 
to the parameters of the constrained minimal
supersymmetric extension of the Standard Model (CMSSM), in which the
SUSY-breaking parameters are universal at the GUT scale, and the
non-universal Higgs model (NUHM), in which this constraint
is relaxed for the soft SUSY-breaking contributions to the Higgs masses.
Within the CMSSM indirect bounds on the mass of the lightest
$\cp$-even Higgs boson are derived.
} 
\maketitle
\section{Introduction}

Phenomenological analyses of supersymmetry (SUSY) often make simplifying
assumptions that drastically reduce the 
dimensionality of the parameter space of the minimal supersymmetric
extension of the Standard Model (MSSM).
One assumption that is frequently employed is
that (at least some of) the soft SUSY-breaking parameters are universal
at some high input scale, before renormalisation. 
One model based on this simplification is the 
constrained MSSM (CMSSM), in which all the soft SUSY-breaking scalar
masses $m_0$ are assumed to be universal at the GUT scale, as are the
soft SUSY-breaking gaugino masses $m_{1/2}$ and trilinear couplings
$A_0$. The assumption that squarks and sleptons with the same gauge
quantum numbers have the same masses is motivated by the absence of
identified supersymmetric contributions to flavour-changing neutral
interactions and rare decays.
Universality between squarks and 
sleptons with different gauge interactions may be motivated by some GUT
scenarios~\cite{GUTs}. 
However, the universality of the soft SUSY-breaking
contributions to the Higgs scalar masses is less motivated, and is
relaxed in the non-universal Higgs model (NUHM)~\cite{NUHM1,NUHM2,NUHMother}. 

In \citere{ehoww}
a combined $\chi^2$~analysis has been performed of electroweak precision
observables (EWPO), going beyond previous such analyses~\cite{ehow3,ehow4} 
(see also \citere{other}), and of $B$-physics observables (BPO),
including some that have not been
included before in comprehensive analyses of the SUSY parameter
space (see, however, \citere{LSPlargeTB}). 
The set of EWPO included in the analysis of \citere{ehoww} are
the $W$~boson mass $\MW$,  the effective leptonic weak mixing angle
$\sweff$, the total $Z$~boson width $\Ga_Z$ 
(using for these three observables the recent theory
predictions obtained in \citeres{MWweber,zobs}), 
the anomalous magnetic moment of the  muon $(g-2)_\mu$
(based on \citeres{g-2FSf,g-2CNH}, see \citere{g-2reviews} for recent
reviews), and the mass of the lightest MSSM Higgs boson $\Mh$
(obtained from the program 
{\tt FeynHiggs}~\cite{feynhiggs,mhiggslong,mhcpv}). 
In addition, four BPO are included:  the branching ratios
$\br(b \to s \ga)$ (based on the results of \citere{bsg},
incorporating also the latest SM corrections provided
in~\citere{bsgtheonew}), $\br(B_s \to \mu^+ \mu^-)$ (based on 
results from \citere{bsmumu},
which are in good agreement with \citere{ourBmumu})
and $\br(B_u \to \tau \nu_\tau)$ (based on \citere{BPOtheo}), 
and the $B_s$ mass mixing parameter $\De M_{B_s}$ (based on
\citere{BPOtheo}). 

For the evaluation of the BPO minimal flavor violation (MFV)
at the electroweak scale is assumed. Non-minimal flavor violation (NMFV) 
effects
can be induced by RGE running from the high scale, see
e.g.\ \citere{nmfv}, that may amount to $\sim 10\%$ of the SUSY
corrections. These additional contributions are neglected in the
present analysis.

For each observable, the $\chi^2$~function is constructed
including both theoretical and experimental systematic uncertainties, as
well as statistical errors~\cite{ehoww}. The analysis is carried out in the 
CMSSM and the NUHM, taking into account 
the fact that the cold dark matter density is
known from astrophysics and cosmology with an uncertainty smaller
than $10~\%$~\cite{WMAP}, 
effectively reducing the dimensionality of the parameter
space by one. The combined $\chi^2$~function for the EWPO and the BPO is
investigated in the CMSSM and the NUHM. For the CMSSM furthermore indirect
constraints on the lightest Higgs-boson mass, $\Mh$, are discussed.


\section{CMSSM analysis including EWPO and BPO}
\label{sec:cmssm}

\begin{figure}[tbh!]
\begin{center}
\includegraphics[width=.48\textwidth]{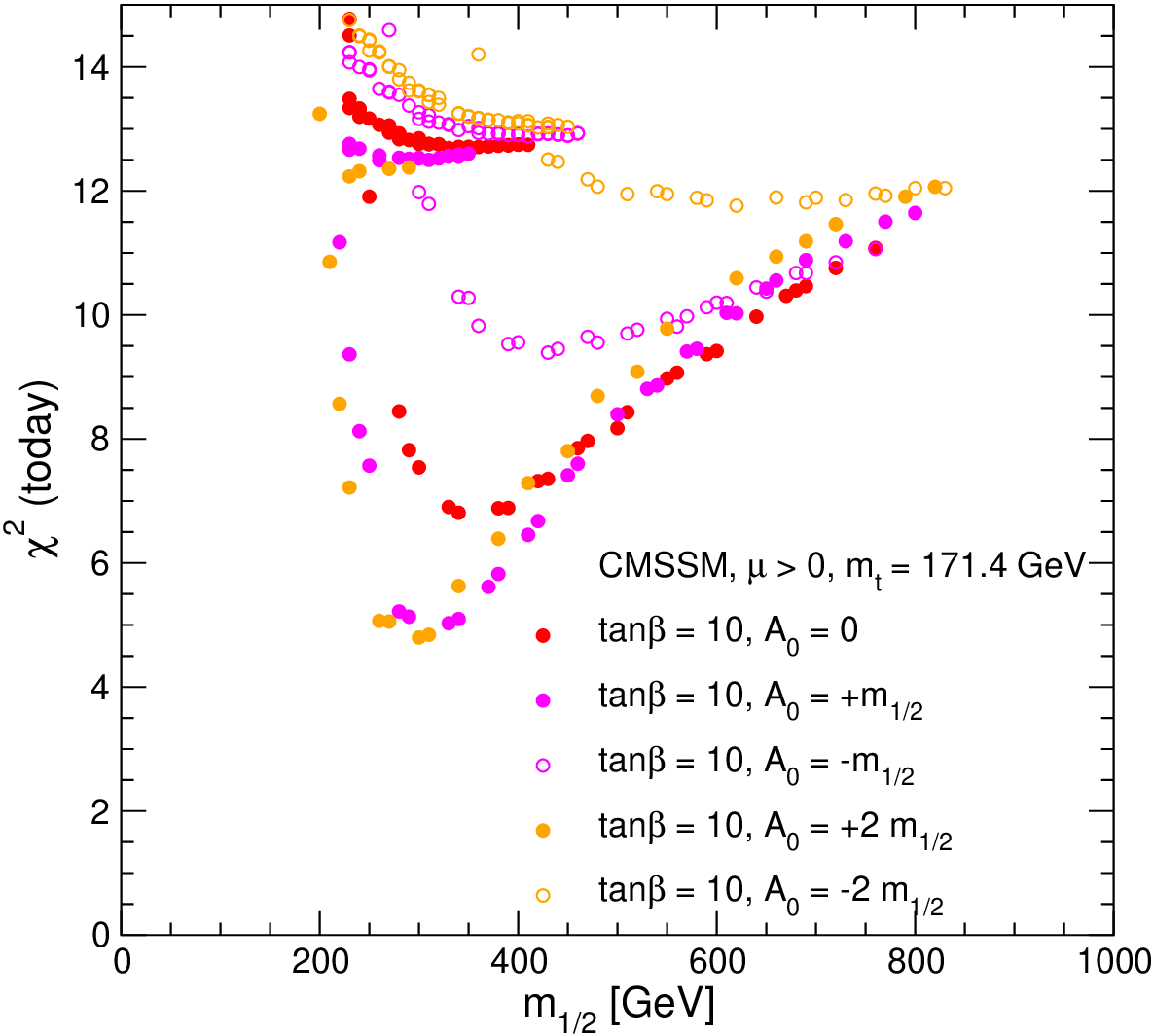}\\
\includegraphics[width=.48\textwidth]{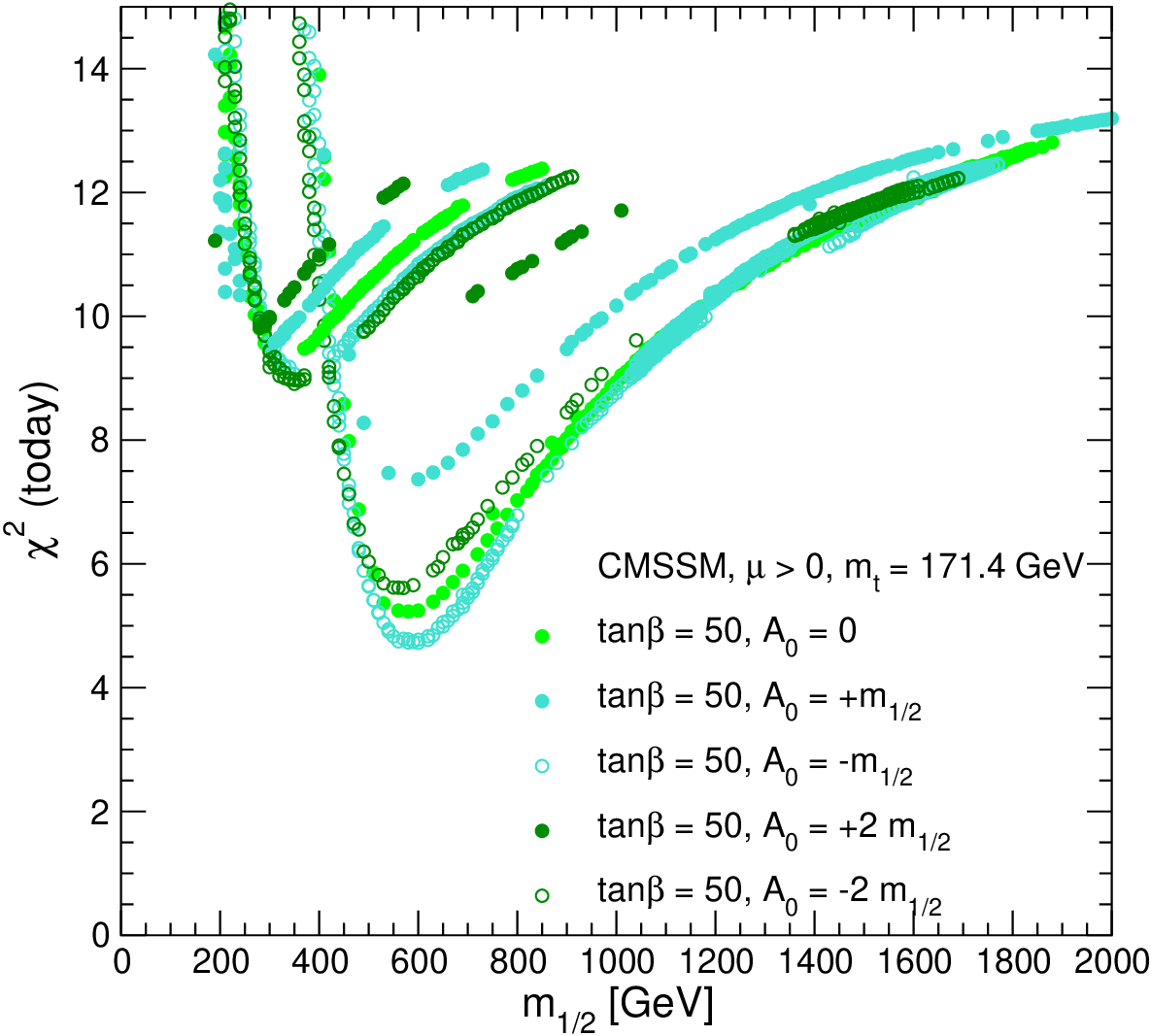}
\caption{%
The combined $\chi^2$~function for the electroweak
observables $\MW$, $\sweff$, $\Ga_Z$, $(g - 2)_\mu$, $\Mh$, 
and the $b$~physics observables
$\br(b \to s \ga)$, $\br(B_s \to \mu^+\mu^-)$, $\br(B_u \to \tau \nu_\tau)$
and $\De M_{B_s}$, evaluated in the CMSSM for $\tb = 10$ (upper plot) and
$\tb = 50$ (lower plot) for various discrete values of $A_0$. We use
$\mt = 171.4 \pm 2.1 \gev$ and $\mb(\mb) = 4.25 \pm 0.11 \gev$, and 
$m_0$ is chosen to yield the central value of the cold dark matter
density indicated by WMAP and other observations for the central values
of $\mt$ and $\mb(\mb)$.}
\label{fig:chi}
\end{center}
\end{figure}

In \reffi{fig:chi} we show for the CMSSM the combined $\chi^2$~values
for the EWPO and BPO, computed as described in \citere{ehoww}, for 
$\tb = 10$ (upper panel) and $\tb = 50$ (lower panel).
We see that the global minimum of $\chi^2 \sim 4.5$
for both values of $\tb$. This is quite a good fit for the number of
experimental observables being fitted. There is a slight tension between
the EWPO, which show a preference for small $m_{1/2}$, and the BPO,
which do not exhibit this behaviour, see \citere{ehoww} for a more
detailed discussion.
For both values of $\tb$, the focus-point region is disfavoured by comparison
with the coannihilation region, though this effect is less important for
$\tb = 50$. 
For $\tb = 10$, $m_{1/2} \sim 300 \gev$ and $A_0 > 0$ are preferred, whereas,
for $\tb = 50$, $m_{1/2} \sim 600 \gev$ is preferred, and there is a
slight preference for $A_0 < 0$. This
change-over is largely due to the impact of the LEP $\Mh$ constraint for
$\tb = 10$ and the $b \to s \ga$ constraint for $\tb = 50$.

\begin{figure}[tbh!]
\begin{center}
\includegraphics[width=.48\textwidth]{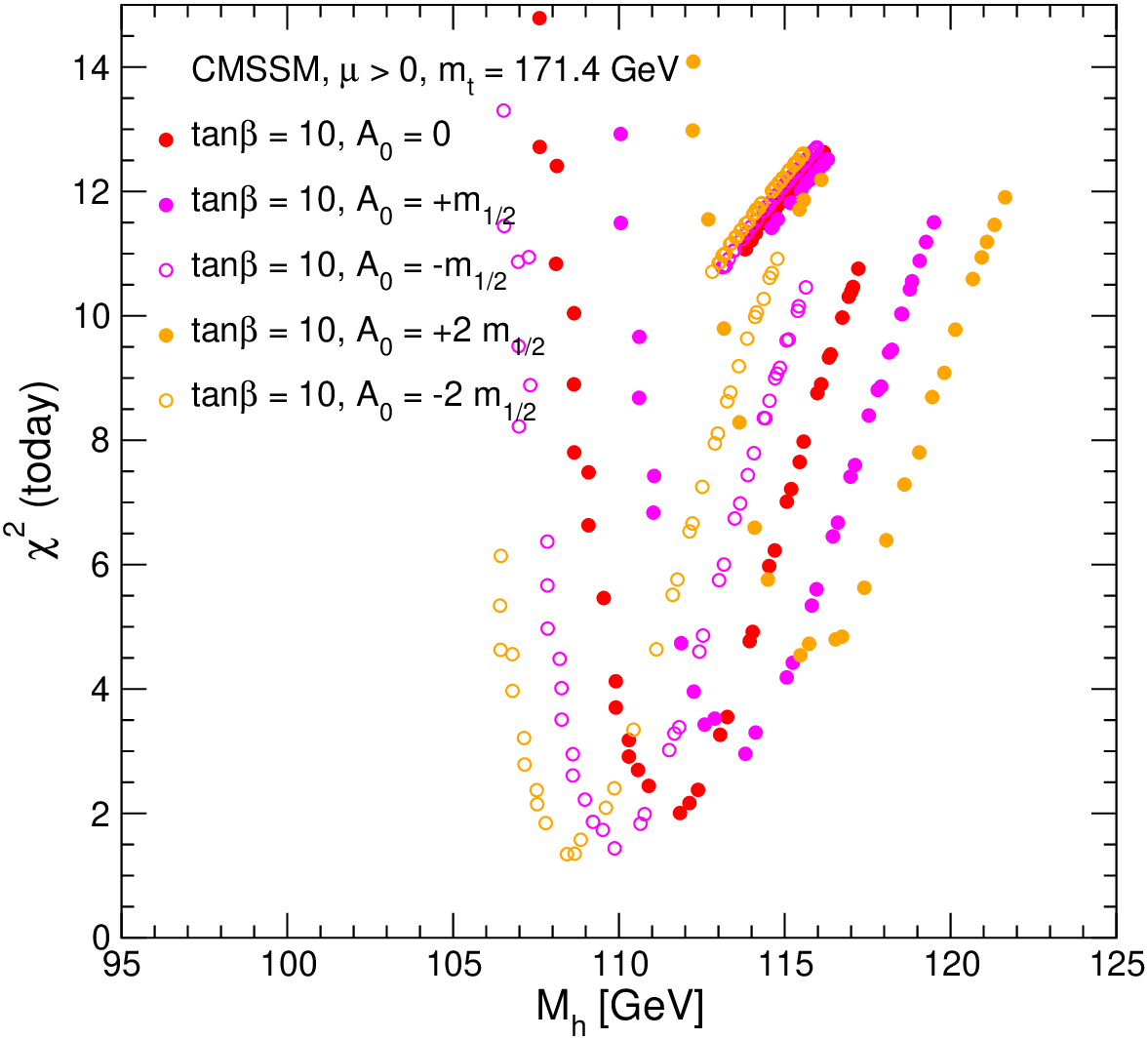}
\includegraphics[width=.48\textwidth]{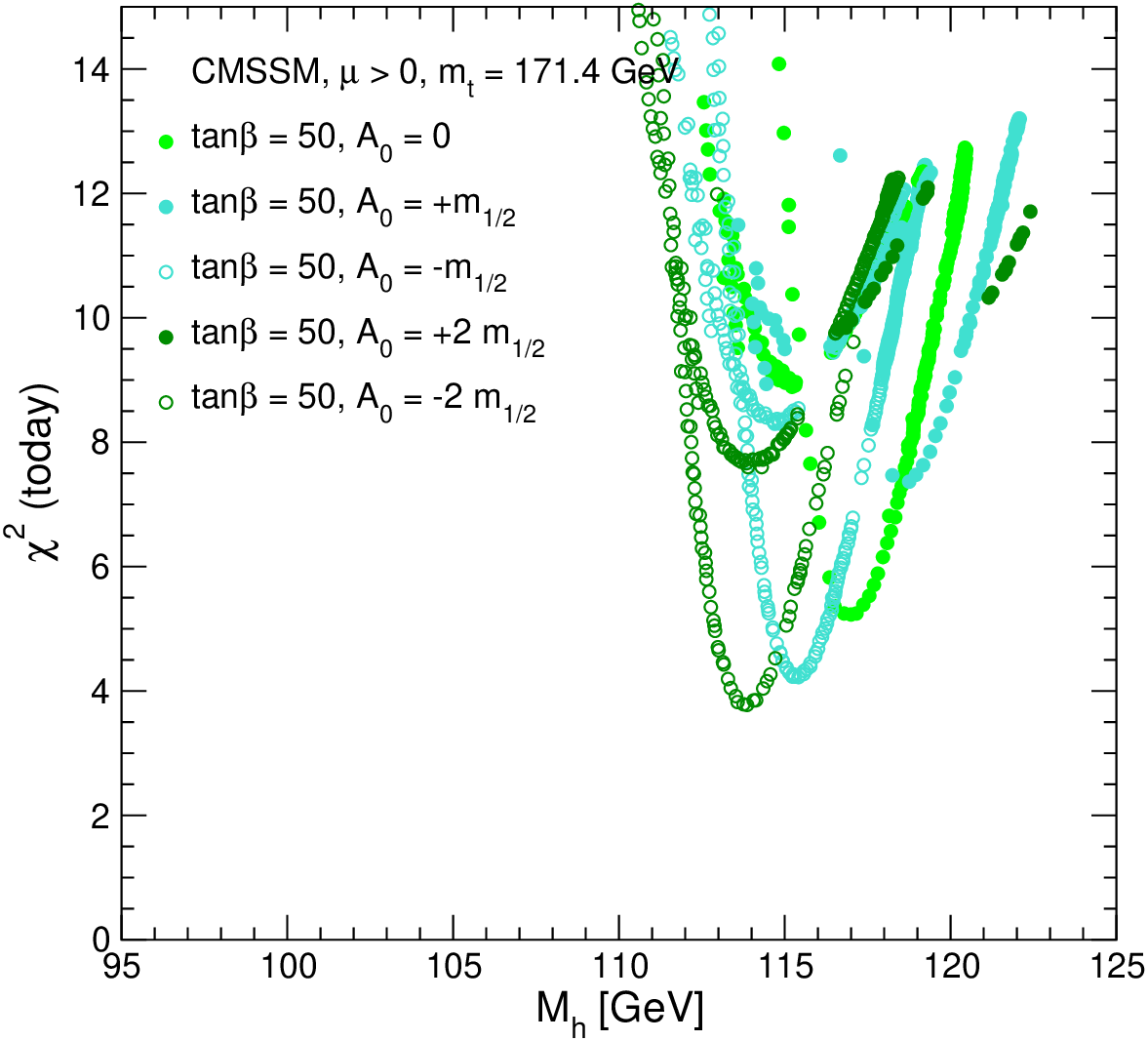}
\caption{%
The combined $\chi^2$~function for $\Mh$, as obtained from a
combined analysis of all EWPO and BPO {\it except\/} the LEP Higgs search,
as evaluated in the CMSSM for $\tb = 10$ (upper plot) and
$\tb = 50$ (lower plot) for various discrete values of $A_0$. We use
$\mt = 171.4 \pm 2.1 \gev$ and $\mb(\mb) = 4.25 \pm 0.11 \gev$, and 
$m_0$ is chosen to yield the central value of the cold dark matter
density indicated by WMAP and other observations for the central values
of $\mt$ and $\mb(\mb)$.}
\label{fig:noLEPMh}
\end{center}
\vspace{-1em}
\end{figure}

In \reffi{fig:noLEPMh}
we display the total
$\chi^2$~functions for $\Mh$, 
as calculated in the CMSSM for $\tb = 10$ (upper panel) and $\tb = 50$
(lower panel) including the information from all EWPO and BPO, 
{\em except\/} from the direct Higgs search at LEP. 
This corresponds to the fitted value of $\Mh$ in the CMSSM. 
In the case of the SM, it is well known that tension between the
lower limit on $\Mh$ from the LEP direct search and the relatively low
value of $\Mh$ preferred by the EWPO has recently been
increasing~\cite{LEPEWWG,TEVEWWG}. 
\reffi{fig:noLEPMh} shows that this tension is significantly
reduced within 
the CMSSM, particularly for $\tb = 50$.  
We see that all data (excluding $\Mh$) favour a
 value of $\Mh \sim 110 \gev$ if $\tb = 10$ and $\Mh \sim 115 \gev$ if
$\tb = 50$. On the other hand, 
the currently best-fit value for the SM Higgs boson
of $\MHSM$ is $76 \gev$~\cite{LEPEWWG},
i.e.\ substantially below the SM LEP bound of $114.4 \gev$~\cite{LEPHiggsSM}.
Our results for the indirect constraints on $\Mh$ have meanwhile 
been confirmed by a more elaborate $\chi^2$
fit where all CMSSM parameters and the constraint from the dark matter 
relic density are included in the fit~\cite{mastercode}.

In \citere{ehoww} we have also determined the total $\chi^2$~functions for 
$\Mh$ based on the information from all EWPO and BPO, {\em including\/} 
the limit from the direct Higgs search at LEP. In this case the favoured 
$\Mh$ value for $\tb = 10$ is increased by $\sim 5 \gev$,
whereas the difference is only $\sim 1 \gev$ if $\tb = 50$.

\section{NUHM analysis including EWPO and BPO}
\label{sec:nuhm}

\begin{figure}[tbh!]
\begin{center}
\includegraphics[width=.40\textwidth]{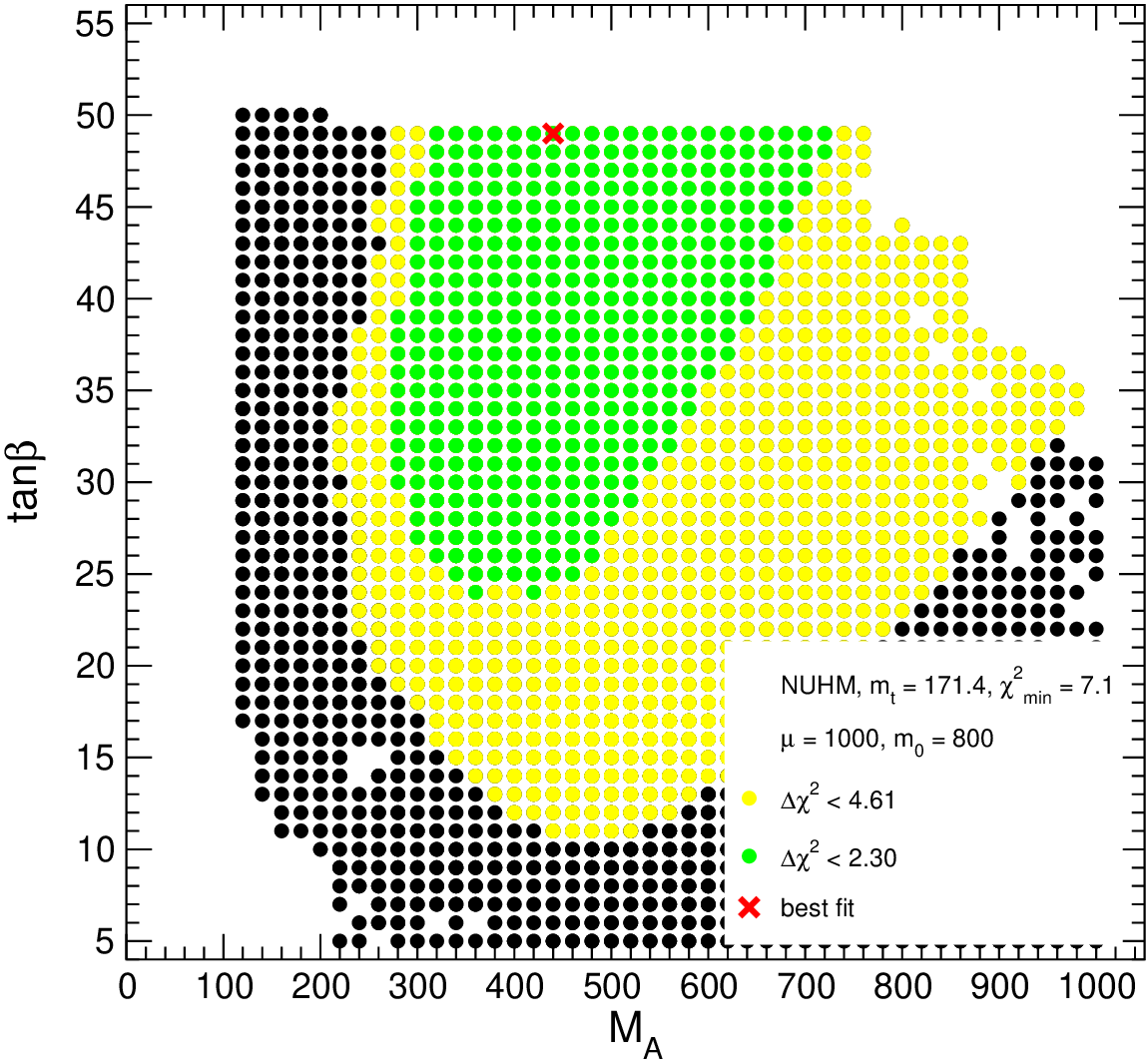}
\end{center}
\caption{%
The combined EWPO and BPO $\chi^2$~function
for a WMAP-compatible 
\plane{\MA}{\tb}\ 
in the  NUHM (plane 
\Athree\ of \citeres{ehoww,ehhow}). 
We use $\mt = 171.4 \pm
2.1 \gev$ and $\mb(\mb) = 4.25 \pm 0.11 \gev$, and $m_{1/2}$ is adjusted
continuously so as to yield the central value of the cold dark matter
density indicated by WMAP and other observations for the central values
of $\mt$ and $\mb(\mb)$.}
\label{fig:A3}
\end{figure}

The NUHM has two more parameters in addition to those of the CMSSM.
They characterise the degree of non-universality of the two Higgs mass
parameters. After imposing the electroweak vacuum conditions
the two parameters can be traded for $\MA$ and $\mu$. It has been
pointed out in \citeres{ehoww,ehhow} that 
$m_{1/2}$ or $\mu$ can be varied such that (essentially) the whole
\plane{\MA}{\tb} 
is compatible with the WMAP constraint on the dark matter relic density.

\reffi{fig:A3} shows the combined EWPO and BPO $\chi^2$~function
for a \plane{\MA}{\tb} in the NUHM
(called plane \Athree\ in \citeres{ehoww,ehhow}) 
with $m_0 = 800 \gev$ and 
$\mu = 1000 \gev$, where $m_{1/2}$ is chosen to vary across the plane so as
to maintain the WMAP
relationship with $\MA$:
\begin{equation}
\frac{9}{8} \MA  - 12.5  \gev \le m_{1/2} \;  \le \frac{9}{8} \MA +
37.5\gev.
\label{A3}
\end{equation}
The best-fit point in this example has $\MA \sim 440 \gev$
and $\tb \sim 50$. 
It has $\chi^2 = 7.1$, which is 
slightly worse than the CMSSM fits
in \reffi{fig:chi}.
We also display the
$\De \chi^2 = 2.30$ and 4.61 contours, which would correspond to the
68~\% and 95~\% C.L. contours in the \plane{\MA}{\tb} {\it if\/} the
overall likelihood distribution, $\cL \propto e^{-\chi^2/2}$,
was Gaussian. This is clearly only roughly the case in this
analysis, but these contours nevertheless give interesting indications
on the preferred region in the \plane{\MA}{\tb}.
No results are shown in the upper right corner of the plane
(with high $\MA$ and high $\tb$) because there the relic density 
is low compared to the preferred WMAP value.
The lower left portion of
the plane is missing because of the finite resolution of our
scan. 

\subsection*{Acknowledgements}

The author thanks J.~Ellis, S.~Heinemeyer, K.~Olive and A.~Weber
for collaboration on the results presented in this paper.
Work supported in part by the European Community's Marie-Curie Research
Training Network under contract MRTN-CT-2006-035505
`Tools and Precision Calculations for Physics Discoveries at Colliders'.

\end{document}